\documentclass[16pt,superscriptaddress,showpacs,preprintnumbers,amsmath,amssymb,floatfix,prd]{revtex4}

\usepackage{graphicx}
\usepackage{dcolumn}
\usepackage{bm}
\usepackage{hyperref}
\usepackage{url}
\usepackage{amsfonts}
\usepackage{amsmath}
\hypersetup{colorlinks,citecolor=blue}

\begin{document}

\title{Evolving wormhole geometry from dark matter energy density}

\author{Farook Rahaman}
\email{rahaman@associates.iucaa.in}
\affiliation{Department of Mathematics, Jadavpur University, Kolkata 700032, West Bengal, India}

\author{Bikramarka S Choudhury}
\email[Email:]{ bikramarka@gmail.com}
\affiliation{Department of Mathematics, Jadavpur University, Kolkata 700032, West Bengal, India}

\date{\today}

\begin{abstract}

We analyse  traversable wormholes  defined by the dynamic line elements that asymptotically approaches Friedmann-Robertson-Walker (FRW) universe. This dynamical wormholes is supported by the galactic dark matter as well as perfect isotropic fluid. We will discuss several evolving Lorentzian wormholes comprising with different perfect isotropic fluids in addition to various scale factors. We will speculate the various significance, features and throat energy conditions for these evolving traversable  Lorentzian wormholes.

\end{abstract}

\maketitle

\section{Introduction:}
According to Einstein's theory \cite{PartilcleProblem} ,  the spacetime    fabric and geometry
due to  the presence of matter is not inflexible but is adaptable and deformable.
Heavier condensed object  produces  strong the curvature of space \cite{003},  which fundamentally leads to the concepts of black holes \cite{001,002} containing  the curvature singularity.  Due to the presence of singularity, one can not travel through it, however, in some way, if a structure could be constructed  that consists of a throat then it might be possible to travel through the structure. Fortunately, Einstein field equations \cite{einstein1915field} may have such solutions under certain conditions where the spacetime geometry is distorted such a manner that creates a tunnel like structure in the galactic fabric. These theoretical solutions, initially proposed by Albert Einstein and Nathan Rosen, in the year 1935, are known as Einstein-Rosen bridges or Wormholes \cite{FRWwormhole, Wormhole_collapse}. This structure connects two different points in the spacetime fabric in a way that is analogous to a shortcut that can ultimately minimize travel time and make it seem like the travel speed is even greater than that of light. But to be able to travel through a wormhole, certain conditions for the wormhole must be satisfied, for example the throat of the wormhole structure should be stable enough for enabling one to travel through it \cite{traversable}, also the wormhole must have a structure that allows the size of the traveler's body to pass through it. Wormholes that satisfy such conditions are called traversable wormholes \cite{Travworm_state,CosmoModel}. We understand that the throat of the wormhole is one essential thing for the structure and a necessary condition for construction of throat is presence of exotic matter which violates the Null-Energy condition \cite{004} . Now a big concern is whether such matter exists. In answer to that we would like to point to the recent finding of the accelerated expansion of the universe \cite{observational}. In various attempts to explain this phenomenon, the arguments have led to the possible existence of matter with a large negative pressure. These are referred to as dark matter or dark energy \cite{cosmology,observational}. The dark matter or dark energy, which are basically exotic matter, play a pivotal role in the formation of wormhole structure \cite{Wormhole_darkEnergyVladimir}. Now, since the universe is expanding and wormholes are structures inside the geometry of the 4-dimensional spacetime manifold, so the idea is that with the expansion of universe, there is a corresponding effect on the geometry of the manifold as the universe is expanding from every point and hence it must have an effect on the wormhole structure as well. The wormholes that exhibit this property are called evolving wormholes \cite{Revisiting_perspec,Phatom_singularities,EvoLorentzian, EvolvingTopologically} i.e. the wormholes are evolving with time. We know that the scale factor is a function of time. So we have attempted to determine corresponding changes in the wormhole structures as the time varies.  

In this paper we have discussed several aspects of the evolving wormholes. In the former sections, we briefly describe metric and Einstein field equations regarding wormhole in FRW universe. Then we dive into solutions regarding wormholes as well as scale factors and density functions for different cases. After that we study the energy conditions and embedding space regarding the wormhole and next, we calculate the condition for traversability through the wormhole and finally we calculate the proper length between two distances.

 \section{ basic equations of WORMHOLE EMBEDDED IN frw universe:}

 Let us  consider the spacetime metric of
a dynamic traversable wormhole in a Friedmann-Robertson-Walker universe as

\begin{equation}
    ds^2 = - e^{2 \phi (r)} dt^2 + a^2(t) \left[\frac{dr^2}{1 - k r^2 - \frac{b(r)}{r} } + r^2 d \theta ^2 + r^2 \sin ^2 \theta d \phi ^2\right].
\end{equation}
  Here  b(r) and $\phi(r)$   are known as  shape function ( describes the nature of the wormhole throat which  indicates the surface of minimum radius ) and redshift function ( which is finite everywhere to ensure  without event horizon ) respectively and    $ a(t) $ is the scale factor of the universe describing  the size of the universe.
$k$  represents the sign of the curvature of spacetime with values:
$+1, 0, -1$.  Note that the   metric (1)  becomes static Morris-Thorne wormhole  when   $ a(t) \rightarrow ~ constant  $ and $ k \rightarrow ~ 0  $.
When  $ b(r) ~and ~\phi (r) \rightarrow  0$, the
  the spacetime metric (1)  coincides with  the FRW metric.

In this study, we assume an  inhomogeneous  and  anisotropic  fluid  matter source comprising the wormhole with a diagonal energy-momentum tensor   as $ (\rho(r, t), p_r(r, t), p_t(r, t), p_t(r, t))$. The quantities  $ \rho(r, t), p_r(r, t), p_t(r, t)$ stand for the mass energy density, radial
pressure  and  transverse
pressure  as  measured  by  observers  sited  at            constant
$r,u , \phi$.

   With these  considerations,
the non trivial Einstein's equations are obtained as ( $ 8 \pi G  = 1$ )

\begin{equation}
    \rho (r,t) = \frac{3 \dot a ^2}{a^2} e^{-2 \phi} + \frac{3k}{a^2} + \frac{b'}{a^2 r^2},
\end{equation}
\begin{equation}
  p_r (r,t) = - \frac{2 \ddot a }{a} e^{-2 \phi} - \frac{ \dot a ^2}{a^2} e^{-2 \phi} - \frac{k}{a^2} - \frac{b}{a^2 r^3} + \frac{2}{a^2 r} \phi ' \left(1 - kr^2 - \frac{b}{r}\right),
\end{equation}
\begin{multline}
    p_t (r,t) = - \frac{2 \ddot a }{a} e^{-2 \phi} - \frac{ \dot a ^2}{a^2} e^{-2 \phi} - \frac{k}{a^2} + \frac{b-r b'}{2 a^2 r^3}
     + \frac{1}{a^2 r} \phi ' \left(1 - k r^2 - \frac{b}{r}\right)\\+ \frac{1}{a^2} \left[{\phi '} ^2 \left(1- k r^2 - \frac{b}{r}\right)  + \phi''\left(1 - k r^2 - \frac{b}{r}\right) - \frac{1}{2} \phi'\left(2k r + \frac{r b' -b}{r^2}\right) \right] ,
\end{multline}
\begin{equation} \label{5}
    T_{t r} = \frac{\dot a}{a^2} e^{- \phi}  {\phi ' \left(1 - k r^2 - \frac{b}{r}\right)}^{1/2} .
\end{equation}
 Here, a  prime  and an
overdot denote differentiation with respect to r and t, respectively.   $T_{t r}$ stands  for the outward energy flow.

               Conservation of energy  equation, $T_{\mu ; \nu}^\nu = 0$, yields
\begin{equation}
    \frac{\partial \rho}{ \partial t} + \frac{\dot a}{a} (3 \rho + p_r + 2 p_t) = 0,
\end{equation}
\begin{equation}
    p_r ' - (p_t - p_r) \frac{2}{r} =0.
\end{equation}

\section{ evolving wormhole solutions :}

We have assumed no outward energy flow, i.e. $T_{t r} =0$.

Equation (5)  implies either  $\phi' = 0$ or $\dot a =0$.  If $\dot a =0$, then dynamism of wormhole will be lost. So, we reject this case.
Now, $\phi ' = 0$ implies
$\phi = constant = 0$.\\

Now the field equations (2)-(4) will be modified as
\begin{equation}
    \rho (r, t) = \frac{3(\dot a ^2 +k)}{a^2} + \frac{b'}{a^2 r^2},
\end{equation}
\begin{equation}
    p_r (r,t)= - \frac{2 \ddot a}{a} - \frac{\dot a ^2}{a^2} - \frac{k}{a^2} - \frac{b}{a^2 r^3},
\end{equation}
\begin{equation}
    p_t (r,t)= - \frac{2 \ddot a}{a} - \frac{\dot a ^2}{a^2} - \frac{k}{a^2} + \frac{b - r b'}{2 a^2 r^3}.
\end{equation}

Following  Sung-Won Kim [1], we use the method of separation of variables  to solve these equations, and assume

\begin{align} \label{11}
    \begin{split}
        a^2 (t) \rho(r, t) & = a^2(t) \rho ^c (t) + \rho ^w (r), \\
        a^2 (t) p_r(r, t) & = a^2(t) p^c (t) + p_r ^w (r), \\
        a^2 (t) p_t(r, t) & = a^2(t) p^c (t) + p_t ^w (r).
    \end{split}
\end{align}
 The energy density, radial and transverse  pressures  are taken as separable forms with superscripts $c$ and $w$  indicating  the cosmological part  ( function of t only ) and  the wormhole part  ( function of r only ) respectively. The  cosmological part is characterized by an isotropic pressure $p^c $.

Now, the field equations (8) - (10)  will be separated as

\begin{equation}
    a^2 \left[\rho^c -  \frac{3 (\dot a ^2 +k )}{a^2}\right] = \frac{b'}{r^2} - \rho^w = L,
\end{equation}
\begin{equation}
    a^2\left[p^c + \frac{2 \ddot a}{a} + \frac{\dot a ^2}{a^2} + \frac{k}{a^2}\right] = - \frac{b}{r^3} - p_r ^w = M,
\end{equation}
\begin{equation}
     a^2\left[p^c + \frac{2 \ddot a}{a} + \frac{\dot a ^2}{a^2} + \frac{k}{a^2}\right] = - \frac{b- r b'}{2 r^3} - p_t ^w = M,
\end{equation}
where L  and M are separation constants. In Equations   (13) and
(14), the cosmological parts are equal and that is  why
 the separation constants are the same. Here   both the separation constants should not  be  zero simultaneously as it play the role of connection    between  the  cosmological  part  and  wormhole  part.

Cosmological part, equations (13)  $\&$ (14) yield
\begin{equation}
    \dot \rho _c + \frac{3 \dot a}{a} ( \rho_c + p_c) = (L +3M) \frac{\dot a}{a^3}.
\end{equation}
[ One can also obtain the above equation from conservation of energy  equation, $T_{\mu ; \nu} ^\nu = 0$ ]\\

From equation (12), we can get the cosmological part as
\begin{equation} \label{16}
    \rho_c = \frac{L}{a^2} + \frac{3(\dot a ^2 + k)}{a^2}.
\end{equation}
\\
For cosmological part i.e. evolution part, equation (15) $\&$ (16)  are two master equations with three unknowns namely $a, p_c, \rho_c$.\\
For wormhole part, we have three master equations with four unknowns as
\begin{equation}
    \frac{b'}{r^2} - \rho^w = L,
\end{equation}
\begin{equation}
    \frac{b - b' r}{2r^3} - p_t ^w = M,
\end{equation}
\begin{equation}
    - \frac{b}{r^3} - p_r ^w =M.
\end{equation}

\subsection{   Wormhole solutions supported by dark matter:}

In recent past, Yoshiaki SOFUE \cite{so}  proposed a new density profile of  the dark matter distribution in the spiral galaxy known as the exponential density profile  and is given by
\begin{equation}
    \rho^w = \rho_s e^{-\frac{r}{r_s}},
\end{equation}
where $r_s$ is the  scale radius and $\rho_s$ is the central density.

In the context of evolving wormhole studies, we assume that the
wormhole part is supported by above exponential density profile (20).

Now from  equation (17)   \[  b'= L r^2 + \rho_s  e^{-\frac{r}{r_s}} r^2,\] we obtain
 \begin{equation}
     b = \frac{L}{3} r^3 - \rho_s e^{-\frac{r}{r_s}} [r_s r^2 + 2 r_s ^2 r + 2 r_s ^3] + C,
\end{equation}
where  C is an integration constant. At the throat  radius $r=r_0$, $b(r_0)=r_0$ which implies \[C =r_0 -\frac{L}{3} r_0 ^3 + \rho_s .e^{-\frac{r_0}{r_s}} [r_s r_0 ^2 + 2 r_s ^2 r_0 + 2 r_s ^3].\]

Now, one can find flaring out condition at the throat as $b'(r_0) <1$, which yields
\[Lr_0^2 +r_0^2 \rho_s e^{-\frac{r_0}{r_s}} <1\]
i.e.,
 \begin{equation}L < \frac{1}{r_0^2} - \rho_s e^{-\frac{r_0}{r_s}}.  \end{equation}
The condition for the  asymptotic flatness of the
wormhole is $\lim_{r \rightarrow \infty} \frac{b(r)}{r} \rightarrow 0 $. Therefore, to satisfy the asymptotically flatness condition, one needs to take $L=0$. However, for asymptotically non flat wormhole, one will have to match   the wormhole spacetinme at some junction interface with the Schwarzschild spherically symmetric static vacuum  solution according to Birkhoff theorem.

Putting the value of $b$ in (18) and (19), we can obtain the expressions  of $p_t^w$, $p_r^w$ as
\begin{equation}
    p_t ^w = - \frac{L}{3} - \rho_s e^{-\frac{r}{r_s}}\left[\frac{1}{2} + \frac{r_s}{2r}+\frac{r_s^2}{r^2}+ \frac{r_s^3}{r^3}\right] + \frac{C}{2r^3} - M,
\end{equation}
\begin{equation}
    p_r ^w = - \frac{L}{3} + \rho_s e^{-\frac{r}{r_s}}\left[\frac{r_s}{r} +  \frac{2r_s ^2}{r^2} +   \frac{2r_s ^3}{r^3}\right] - \frac{C}{r^3} - M.
\end{equation}
\\

\subsection{    Solutions for scale factor :}

 Now we will explore the cosmological part with different criteria   as follows:\\


\textbf{Case 1:} Assume power law form of scale factor
\begin{align}
    a(t) & = t^n, ~~~n~is ~an ~arbitrary~ constant.
\end{align}

Here,  we assume  the evolution of cosmological wormhole is described by
  the expansion of the scale factor given in equation (24).

For this scale factor, equations (15) and (16) yield energy density and pressure ( both cosmological part ) as
\begin{equation}
     \rho^c = L t^{-2n} + 3 t^{-2n}(n^2 t^{2n -2} +k),
\end{equation}
\begin{equation}
    p^c = t^{-2n} (M - k) + t^{-2} (2n - 3n^2).
\end{equation}

The total energy density and pressures are given  by
\begin{equation}
   \rho =  L t^{-2n} + 3 t^{-2n}(n^2 t^{2n -2} +k) + t^{-2n}\left[\rho_s e^{-\frac{r}{r_s}}  \right],
\end{equation}
 \begin{equation}
    p_r =  t^{-2n} (M - k) + t^{-2} (2n - 3n^2)+ t^{-2n}\left[- \frac{L}{3} + \rho_s e^{-\frac{r}{r_s}}\left(\frac{r_s}{r} +  \frac{2r_s ^2}{r^2} +  \frac{2r_s ^3}{r^3}\right) - \frac{C}{r^3} - M  \right] ,
\end{equation}
 \begin{equation}
    p_t =  t^{-2n} (M - k) + t^{-2} (2n - 3n^2)+ t^{-2n}\left[- \frac{L}{3} - \rho_s e^{-\frac{r}{r_s}}\left(\frac{1}{2} + \frac{r_s}{2r}+\frac{r_s^2}{r^2}+ \frac{r_s^3}{r^3}\right) + \frac{C}{2r^3} - M  \right].
\end{equation}

\textbf{Case 2:} Assume exponential form of scale factor
\begin{equation}
    a(t) = e^{\omega t}~,~~~\omega~is ~an ~arbitrary~ constant.
\end{equation}
This is known as the de-Sitter universe which  indicates a non singular    continuously expanding model of the universe.

For this choice, we obtain the energy density and  pressure component as
\begin{equation}
    \rho^c = 3 \omega^2 + (L + 3k) e^{-2 \omega t},
\end{equation}
\begin{equation}
   p^c = e^{-2 \omega t}(M - k) - 3 \omega ^2.
\end{equation}
In this case, the total energy density and pressures are given  by
\begin{equation}
   \rho =  3 \omega^2 + (L + 3k) e^{-2 \omega t} +  e^{-2\omega t}\left[\rho_s e^{-\frac{r}{r_s}}  \right],
\end{equation}
 \begin{equation}
    p_r =  e^{-2 \omega t}(M - k) - 3 \omega ^2+ e^{-2\omega t}\left[- \frac{L}{3} + \rho_s e^{-\frac{r}{r_s}}\left(\frac{r_s}{r} + \frac{ 2r_s ^2}{r^2} +  \frac{2r_s ^3}{r^3}\right) - \frac{C}{r^3} - M  \right] ,
\end{equation}
 \begin{equation}
    p_t =   e^{-2 \omega t}(M - k) - 3 \omega ^2+ e^{-2\omega t}\left[- \frac{L}{3} - \rho_s e^{-\frac{r}{r_s}}\left(\frac{1}{2} + \frac{r_s}{2r}+\frac{r_s^2}{r^2}+ \frac{r_s^3}{r^3}\right) + \frac{C}{2r^3} - M  \right].
\end{equation}

\textbf{Case 3:} $\rho ^c = \beta a^{-n_1} (t)$, $\beta, ~ n_1  $ are constants.\\

Here we have assume cosmological density is inversely proportional to polynomial function of the scale factor. With this form of $\rho _c$, equation (16) yields
\[
    \beta a^{2-n_1} = (L+3k) + 3 \dot a ^2. \]
This implies
\begin{equation}
    \int \frac{da}{\sqrt{a^{2-n_1} - (\frac{L+3k}{\beta})}} = t \sqrt{\frac{\beta}{3}} + C_1,
\end{equation}
where, $C_1$ is an integration constant. \\

\textbf{Subcase - 3.1:   $n_1 = 1$ }:
\begin{equation}
    a = \frac{1}{4} \left( t \sqrt{\frac{\beta}{3}} + C_1   \right)^2 + \frac{L+3k}{\beta}
\end{equation}

For this choice, we obtain the energy density and  pressure component as
\begin{equation}
    \rho^c =  \beta \left[\frac{1}{4} \left( t \sqrt{\frac{\beta}{3}} + C_1   \right)^2 + \frac{L+3k}{\beta} \right]^{-1},
\end{equation}
\begin{equation}
   p^c =  (L+3M) \left[\frac{1}{4} \left( t \sqrt{\frac{\beta}{3}} + C_1   \right)^2 + \frac{L+3k}{\beta} \right]^{-2}-\frac{2\beta}{3} \left[\frac{1}{4} \left( t \sqrt{\frac{\beta}{3}} + C_1   \right)^2 + \frac{L+3k}{\beta} \right]^{-1}.
\end{equation}
In this case, the total energy density and pressures are given  by
\begin{equation}
   \rho = \beta \left[\frac{1}{4} \left( t \sqrt{\frac{\beta}{3}} + C_1   \right)^2 + \frac{L+3k}{\beta} \right]^{-1} + \left[\frac{1}{4} \left( t \sqrt{\frac{\beta}{3}} + C_1   \right)^2 + \frac{L+3k}{\beta} \right]^{-2}\left[\rho_s e^{-\frac{r}{r_s}}  \right] ,
\end{equation}
\[
    p_r =  (L+3M) \left[\frac{1}{4} \left( t \sqrt{\frac{\beta}{3}} + C_1   \right)^2 + \frac{L+3k}{\beta} \right]^{-2}-\frac{2\beta}{3} \left[\frac{1}{4} \left( t \sqrt{\frac{\beta}{3}} + C_1   \right)^2 + \frac{L+3k}{\beta} \right]^{-1}  \] \begin{equation} + \left[\frac{1}{4} \left( t \sqrt{\frac{\beta}{3}} + C_1   \right)^2 + \frac{L+3k}{\beta} \right]^{-2}  \left[- \frac{L}{3} + \rho_s e^{-\frac{r}{r_s}}\left(\frac{r_s}{r} +  \frac{2r_s ^2}{r^2} +   \frac{2r_s ^3}{r^3}\right) - \frac{C}{r^3} - M\right] ,
\end{equation}
\[
    p_t = (L+3M) \left[\frac{1}{4} \left( t \sqrt{\frac{\beta}{3}} + C_1   \right)^2 + \frac{L+3k}{\beta} \right]^{-2}-\frac{2\beta}{3} \left[\frac{1}{4} \left( t \sqrt{\frac{\beta}{3}} + C_1   \right)^2 + \frac{L+3k}{\beta} \right]^{-1}\]  \begin{equation} + \left[\frac{1}{4} \left( t \sqrt{\frac{\beta}{3}} + C_1   \right)^2 + \frac{L+3k}{\beta} \right]^{-2} \left[ - \frac{L}{3} - \rho_s e^{-\frac{r}{r_s}}\left(\frac{1}{2} + \frac{r_s}{2r}+\frac{r_s^2}{r^2}+ \frac{r_s^3}{r^3}\right) + \frac{C}{2r^3} - M\right]    .
\end{equation}

\textbf{Subcase - 3.2:   $n_1 = 2$ }:
\begin{equation}
    a = t \sqrt{\frac{1}{3} (\beta - L - 3k)} + C_1
\end{equation}
For this choice, we obtain the energy density and  pressure component as
\begin{equation}
    \rho^c = \beta \left[ t \sqrt{\frac{1}{3} (\beta - L - 3k)} + C_1\right]^{-2} ,
\end{equation}
\begin{equation}
   p^c =\left(l+3M -\frac{\beta}{3}\right) \left[ t \sqrt{\frac{1}{3} (\beta - L - 3k)} + C_1\right]^{-2} .
\end{equation}
In this case, the total energy density and pressures are given  by
\begin{equation}
   \rho =   \beta \left[ t \sqrt{\frac{1}{3} (\beta - L - 3k)} + C_1\right]^{-2}+ \left[ t \sqrt{\frac{1}{3} (\beta - L - 3k)} + C_1\right]^{-2}\left[\rho_s e^{-\frac{r}{r_s}}  \right],
\end{equation}

   \[ p_r = \left(l+3M -\frac{\beta}{3}\right) \left[ t \sqrt{\frac{1}{3} (\beta - L - 3k)} + C_1\right]^{-2} \]\begin{equation}+ \left[ t \sqrt{\frac{1}{3} (\beta - L - 3k)} + C_1\right]^{-2}\left[- \frac{L}{3} + \rho_s e^{-\frac{r}{r_s}}\left(\frac{r_s}{r} +  \frac{2r_s ^2}{r^2} +   \frac{2r_s ^3}{r^3}\right) - \frac{C}{r^3} - M  \right] ,
\end{equation}
\[
    p_t =   \left(l+3M -\frac{\beta}{3}\right) \left[ t \sqrt{\frac{1}{3} (\beta - L - 3k)} + C_1\right]^{-2}\]  \begin{equation} +\left[ t \sqrt{\frac{1}{3} (\beta - L - 3k)} + C_1\right]^{-2}\left[- \frac{L}{3} - \rho_s e^{-\frac{r}{r_s}}\left(\frac{1}{2} + \frac{r_s}{2r}+\frac{r_s^2}{r^2}+ \frac{r_s^3}{r^3}\right) + \frac{C}{2r^3} - M  \right]  .
\end{equation}

\textbf{Case 4: $a(t) = t^{\alpha } exp(\gamma t) $,  $\gamma$ and $\alpha$ are constants:}\\

This is known as   hybrid model of the universe comprising with  power law  form as well as the exponential form of the scale factor.

For this choice, we obtain the energy density and  pressure component as
\begin{equation}
     \rho _c = (L+3k) t^{-2 \alpha } e^{-2 \gamma t} + 3 \left(  \frac{\alpha }{t} + \gamma \right)^2,
\end{equation}
\begin{equation}
    p_c =  (m-k)t^{-2 \alpha } e^{-2 \gamma t} +2 \alpha t^{-2}-3\left(  \frac{\alpha }{t} + \gamma  \right)^2).
\end{equation}
 In this case, the total energy density and pressures are given  by
\begin{equation}
   \rho =   (L+3k) t^{-2 \alpha } e^{-2 \gamma t} + 3 \left(  \frac{\alpha }{t} + \gamma  \right)^2+  t^{-2 \alpha } e^{-2 \gamma t}\left[\rho_s e^{-\frac{r}{r_s}}  \right],
\end{equation}
 \begin{equation}
    p_r =  (m-k)t^{-2 \alpha } e^{-2 \gamma t} +2 \alpha t^{-2}-3\left(  \frac{\alpha }{t} + \gamma  \right)^2) +  t^{-2 \alpha } e^{-2 \gamma t}\left[- \frac{L}{3} + \rho_s e^{-\frac{r}{r_s}}\left(\frac{r_s}{r} +  \frac{2r_s ^2}{r^2} +   \frac{2r_s ^3}{r^3}\right) - \frac{C}{r^3} - M  \right]  ,
\end{equation}
 \begin{equation}
    p_t =  (m-k)t^{-2 \alpha } e^{-2 \gamma t} +2 \alpha t^{-2}-3\left(  \frac{\alpha }{t} + \gamma  \right)^2)+  t^{-2 \alpha } e^{-2 \gamma t}\left[- \frac{L}{3} - \rho_s e^{-\frac{r}{r_s}}\left(\frac{1}{2} + \frac{r_s}{2r}+\frac{r_s^2}{r^2}+ \frac{r_s^3}{r^3}\right) + \frac{C}{2r^3} - M  \right] .
\end{equation}








 The choice of scale factors corresponds to certain case studies. Note that finding solutions to arbitrary scale factors is quite cumbersome. During our case studies we also encountered certain scale factors which we could not solve by existing methods of calculations. Also the choice of the scale factor for the case studies are very standard forms such as exponential form, power law form etc. For these scale factors, we have obtained consistent solutions of the other physical parameters.

\section{embedding space :}

 The radial coordinate 'r' extends from its minimum value at $r_0$, representing the wormhole throat, to infinity, increasing in range.  We possess an asymptotically flat evolving  wormhole characterized by a positive energy density.
 From the metric (1) one can see that   wormholes    at spatial infinity $(r \rightarrow \infty)$  assume the  following asymptotic metric :
\[
    ~~~~~~~~~~~~~~~~~~~~~~~~~~~~~~~~~~~~~~~~~~~~~~~ds^2 = - dt^2 + a^2(t) \left[\frac{dr^2}{1 - k r^2 } + r^2 d \theta^2 + r^2 \sin^2\theta d \phi^2 \right] ~~~~~~~~~~~~~~~~~~~~~~~~~~~~~~~~~~~~~~~~~~(1a)
\]
 The metric exhibits slices where t remains constant, constituting spaces of constant curvature. Consequently, the asymptotic metric (1a) is organized into foliations of spaces with constant curvature. Since the wormhole described by equation (1) evolves over time, each slice at a fixed instant will vary for different time values.

The shape function   actually determines the profile picture of a wormhole.  Consider a slice of the wormhole at a fixed instant of time , $ t= constant = t_0$ and $\theta = \frac{\pi}{2}.$   Here location of $t_0$ within the interval during which the wormhole exists.     Present  wormhole is non-static i.e. it       evolves with  time,  therefore, one can get different   slices  with different   values of time.   As a result,  scale factor $a(t)$ regulates   the shape of the wormhole.
Nevertheless, it can be demonstrated that the wormhole   structure remains unaltered over time through the utilization of an embedding procedure.

The three  dimensional spatial hyper surface    given by $t=t_0$ of our spherically symmetric space-time takes the form
\[
    d \sigma ^2 = g_{rr} d r^2 + r^2 a^2(t_0)(d \theta^2 + \sin ^2 \theta d \phi ^2 ).
\]
It can be embedded in a four dimensional space as
\begin{equation}
    d \sigma ^2 = d \Bar{z}^2 + d \Bar{r} ^2 + \Bar{r} ^2 ( d \theta ^2 + \sin ^2 \theta d \phi ^2 )
\end{equation}
Let us assume, $\Bar{r} = a(t_0) r$, $d \Bar{r} ^2 = a^2(t_0) d r^2$.
\\
In the equatorial place, $\theta =\frac{ \pi }{ 2}$, we have
\begin{align}
    \begin{split}
      d \sigma ^2 & = \frac{a^2(t_0) (t) d r^2}{1 - k r^2 - \frac{b (r)}{r}} + r^2 a^2(t_0) (t) d \theta ^2, \\
      & = d \Bar{z} ^2 + d \Bar{r}^2 + \Bar{r}^2 d \phi ^2, \\
      & = \frac{d \Bar{r}^2}{1 - \frac{\Bar{b}}{\Bar{r}}} + \Bar{r}^2 d \phi ^2.
    \end{split}
\end{align}
 This yields
\begin{equation}
    \frac{d \Bar{z}}{d \Bar{r}} = \pm \left(\frac{\Bar{r}}{\Bar{b}} - 1\right) ^{- \frac{1}{2}} = \pm \left(\frac{a(t_0)r}{a(t_0)b + k a(t_0) r^3} - 1 \right) ^{- \frac{1}{2}} = \frac{dz}{dr},
\end{equation}
\begin{flalign}
    \text{where } && \Bar{b} & = a(t_0)b + k a(t_0) r^3 .&
\end{flalign}
\begin{flalign}
    \text{Note that here } && \Bar{z} & = \pm a(t_0) z &.
\end{flalign}
The flare-out condition for consisting of wormhole i.e. to maintain the shape of the traversable wormhole, the flare-out condition $\frac{d ^2 \Bar{r}}{d \Bar{z} ^2} > 0$ should be satisfied.
\begin{flalign}
    \text{Now, } && \frac{d^2 \Bar{r}}{d \Bar{z} ^2} & = \frac{\Bar{b} - \Bar{b}' \Bar{r}}{2 \Bar{b} ^2}  = \frac{1}{a(t_0)} \frac{b - b' r - 2 k r^3}{2(b + k r^3) ^2} = \frac{1}{a(t_0)} \frac{d^2 r}{d z^2} >0 &.
\end{flalign}
 [ Here, $\Bar{b}'= \frac{d \Bar{b}}{d \Bar{r}}$  ].

Note that for $ k = 0 $, $a(t_0) = 1$, this flare-out condition reduces to flare-out condition of static wormhole. Also one can notice that the evolving wormhole
  will continue the identical size in the $  \Bar{z},  \Bar{r}, \phi$ coordinates. We can draw the graph of the embedded curve $z=z(r)$ as well as the entire imagining of the surface generated  by
the rotation of the embedded curve about the vertical z axis ( figure - 1).

 \begin{figure} [thbp]
\centering
	\includegraphics[width=6cm]{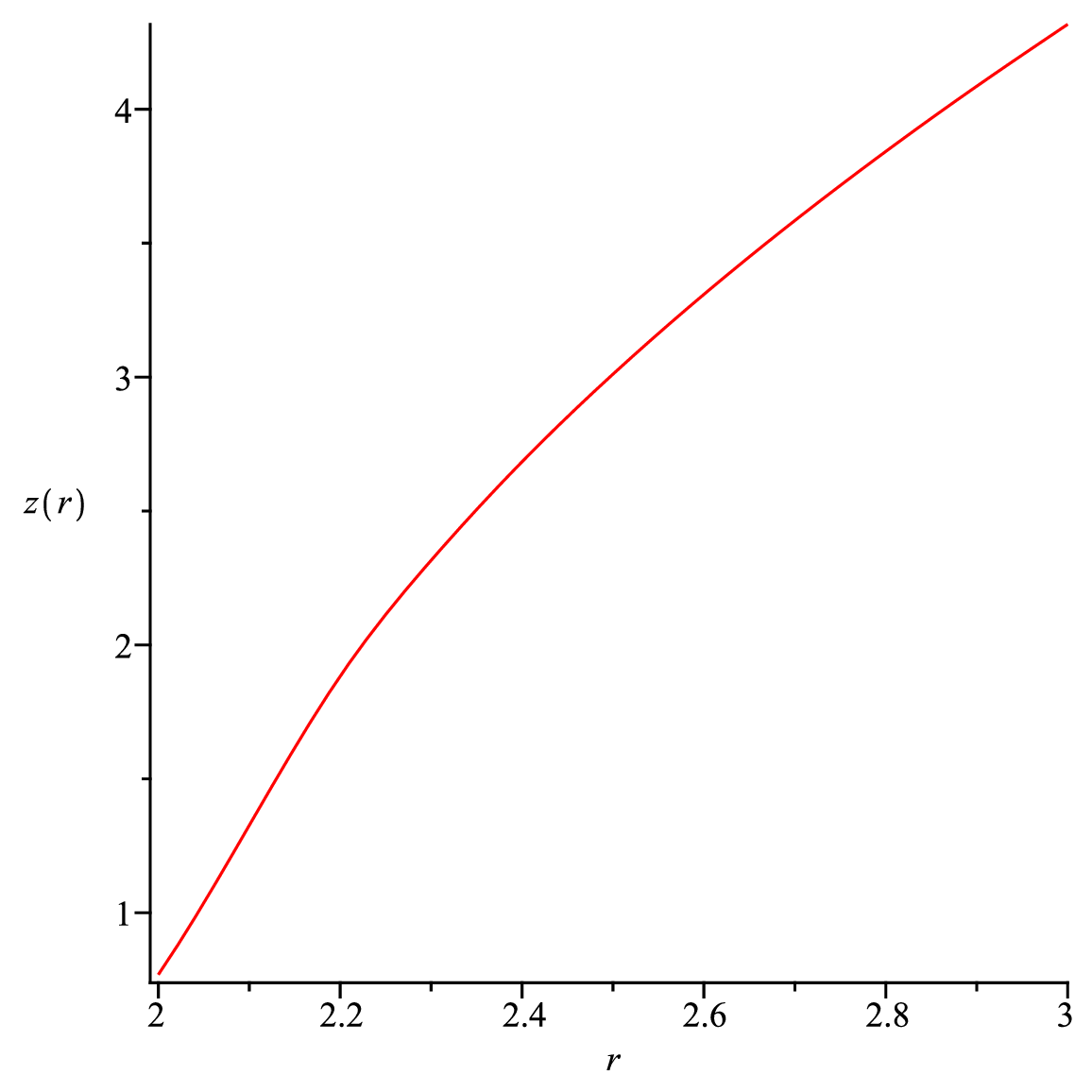}
\includegraphics[width=6.0cm]{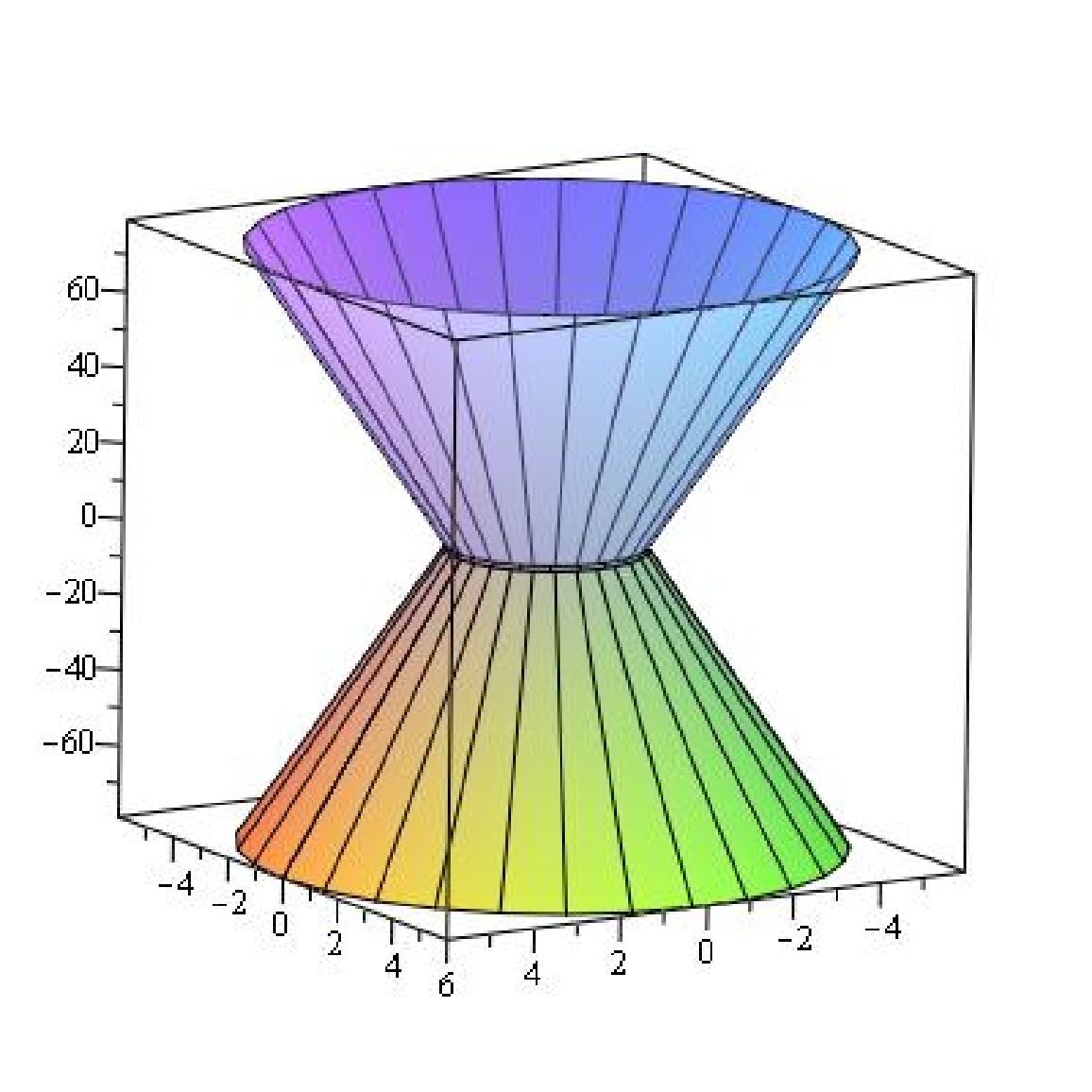}
	\caption {Left panel)  The embedding diagram for the wormhole spacetime given in equation (12). (Right panel) The entire imagining of the surface generated  by
the rotation of the embedded curve about the vertical z axis.}
\end{figure}

 \section{     ENERGY CONDITIONS:}

We have already mentioned that the flaring-out condition has been satisfied ($b'(r_0) <1$). Now at the same time we will try to check whether  null energy condition (NEC) is obeyed or not for constructing the dynamical wormhole.  In static  wormhole configuration, one of the fundamental criterion  is that  the null energy condition should be violated.  However in dynamical wormhole configuration
  the situation may  change due to extra terms (time dependent scale factor) in the field equations. In general one can  execute the condition   $\rho+p_r >0$ for normal matter
comprising the wormhole.   In the event of dynamical wormholes it is argued that that there exist wormhole
solutions which obey NEC.  So, we will search   some
Lorentzian dynamical wormhole geometries  with no requirement of the  matters that violate the NEC. Now, we calculate null energy condition expression as

\begin{align}
    \begin{split}
        \rho + p_r & = 3 \frac{\dot a ^2}{a^2} + 3\frac{k}{a^2} - 2 \frac{\ddot a}{a} - \frac{\dot a ^2}{a^2} - \frac{k}{a^2} - \frac{b}{a^2 r^3} + \frac{b'}{a^2 r^2} \\ & = 2 \frac{\dot a ^2}{a ^2} - 2 \frac{\ddot a}{a} - \frac{(-2k r^3 - r b' +b)}{a^2 r^3} \\ & = A - \left[ \frac{(b+k r^3) ^2}{a r^3} \right] \frac{d^2 \Bar{r}}{d \Bar{z} ^2},
    \end{split}
\end{align}
\begin{flalign}
    \text{where, }  && A & = 2 \frac{\dot a ^2}{a^2} - 2 \frac{\ddot a}{a} &.
\end{flalign}
\\
Since $\frac{d^2 \Bar{r}}{d \Bar{z}^2} > 0$, so sign of $\rho + p_r$ depends on $A$.

For static case $A = 0$, so one gets $\rho + p_r$ always negative i.e. matter distribution should be exotic in nature.

In usual FRW cosmological model of the universe, one can get different scale factor $a(t)$ which depends on $k$ and some parameters.

If $A>0$, then there exists a possibility to have non exotic matter $(p_r + \rho > 0)$ comprising the wormhole.

Here, we calculate  A   for all cases with subcases.

\begin{equation}\text{Case 1: } A = \frac{2n}{t^2}, \end{equation}
\begin{equation}\text{Case 2: }  A = 0 ,\end{equation}
\begin{equation}\text{Case 3.1: }   A = \frac{\frac{\beta}{12}\left(\sqrt{\frac{\beta}{3}}t +C_1\right)^2 -\frac{L+3k}{3} }{\left[\frac{1}{4}\left(\sqrt{\frac{\beta}{3}}t +C_1\right)^2+ \frac{L+3k}{\beta}\right]^2},\end{equation}
\begin{equation}\text{Case 3.2: }  A  = {\frac {2 \beta-2L-6k}{ \left( t\sqrt {3\beta-3L-9k}+\sqrt{3}{
  C_1} \right) ^{2}}},
\end{equation}
 \begin{equation}\text{Case 4: }  A = \frac{2 \alpha }{t^2} .\end{equation}
 Note that $A > 0$ for positive values of the parameters $ n$ , $ \alpha$ and $ \beta> L+3k$.

Now we check whether the NEC is violated or not at the throat $r = r_0$   for all t.
\[(\rho + p_r)_{r=r_0} = 2 \frac{\dot a ^2}{a ^2} - 2 \frac{\ddot a}{a} - \frac{(-2k r_0^3 - r_0 b'(r_0) +b(r_0))}{a^2 r_0^3}. \]
For $(\rho + p_r)_{r=r_0} >0 $,  we have ( using $b(r_0) = r_0$) ,
\begin{equation}
    \dot a ^2 -a \ddot a > \frac{\delta -kr_0^2 }{r_0^2},
\end{equation}
where $2 \delta = [1-b'(r_0)] > 0$, since $b'(r_0)<1$.
\begin{figure} [thbp]
\centering
	\includegraphics[width=5cm]{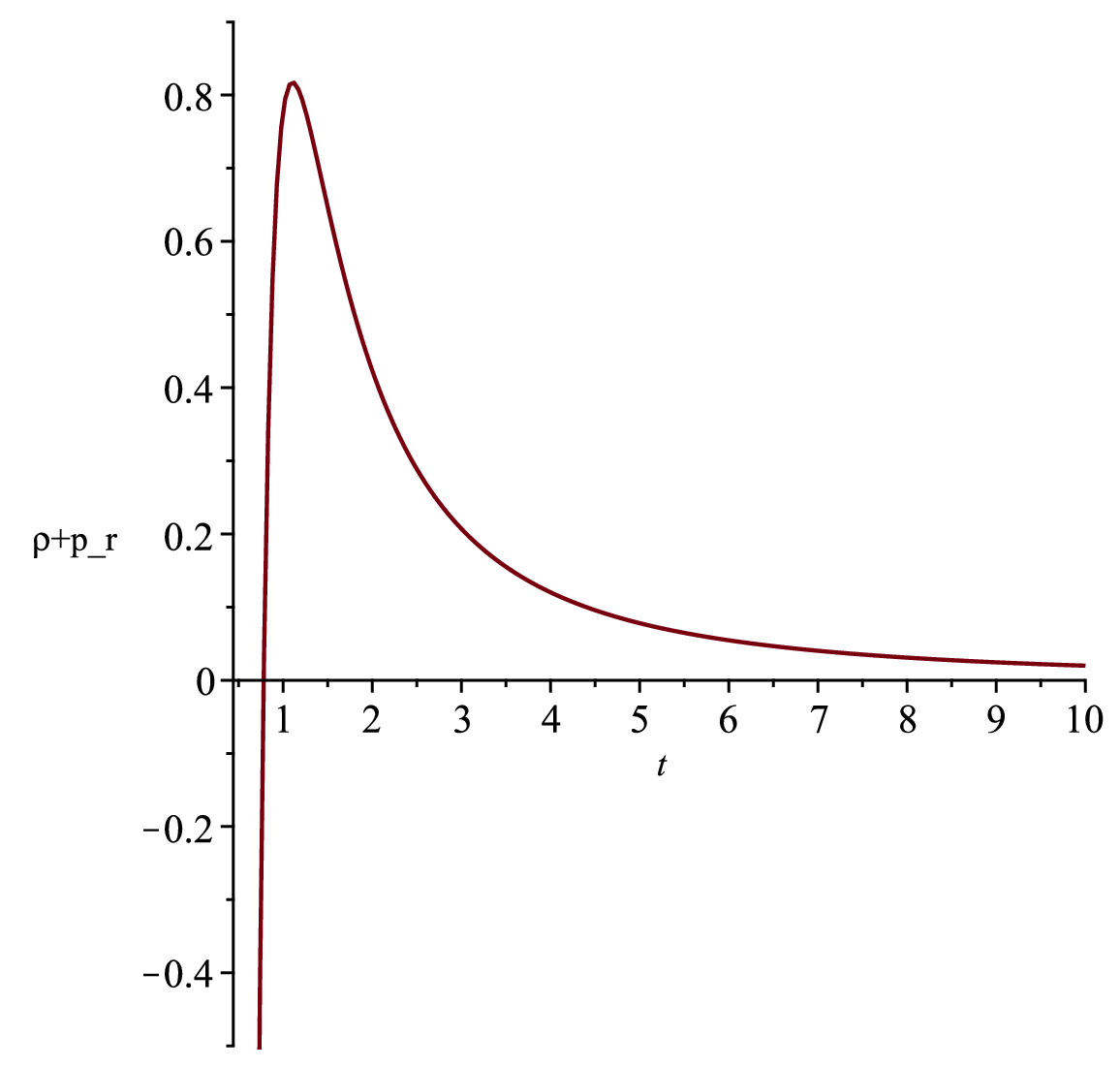}
\includegraphics[width=5cm]{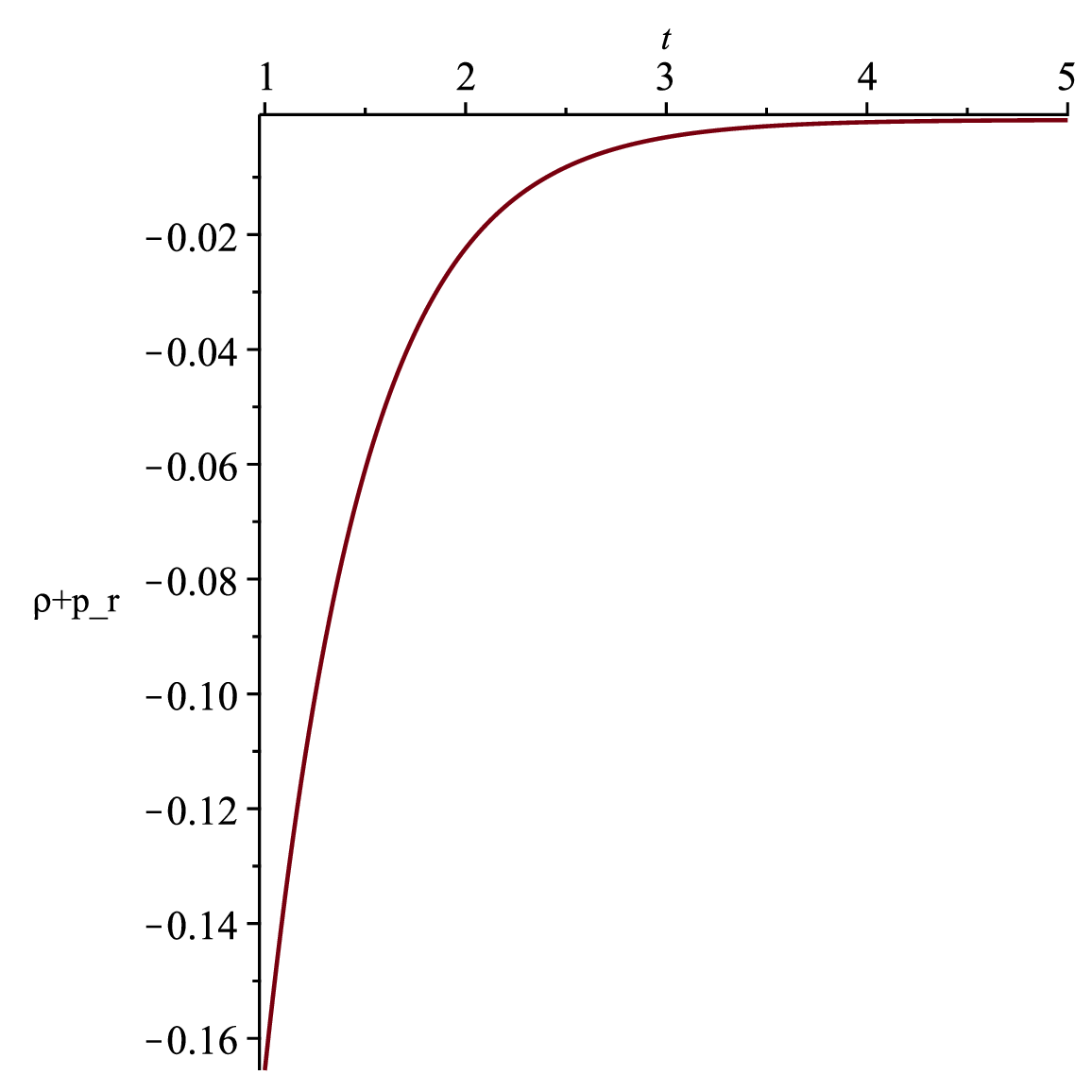}
	\includegraphics[width=5cm]{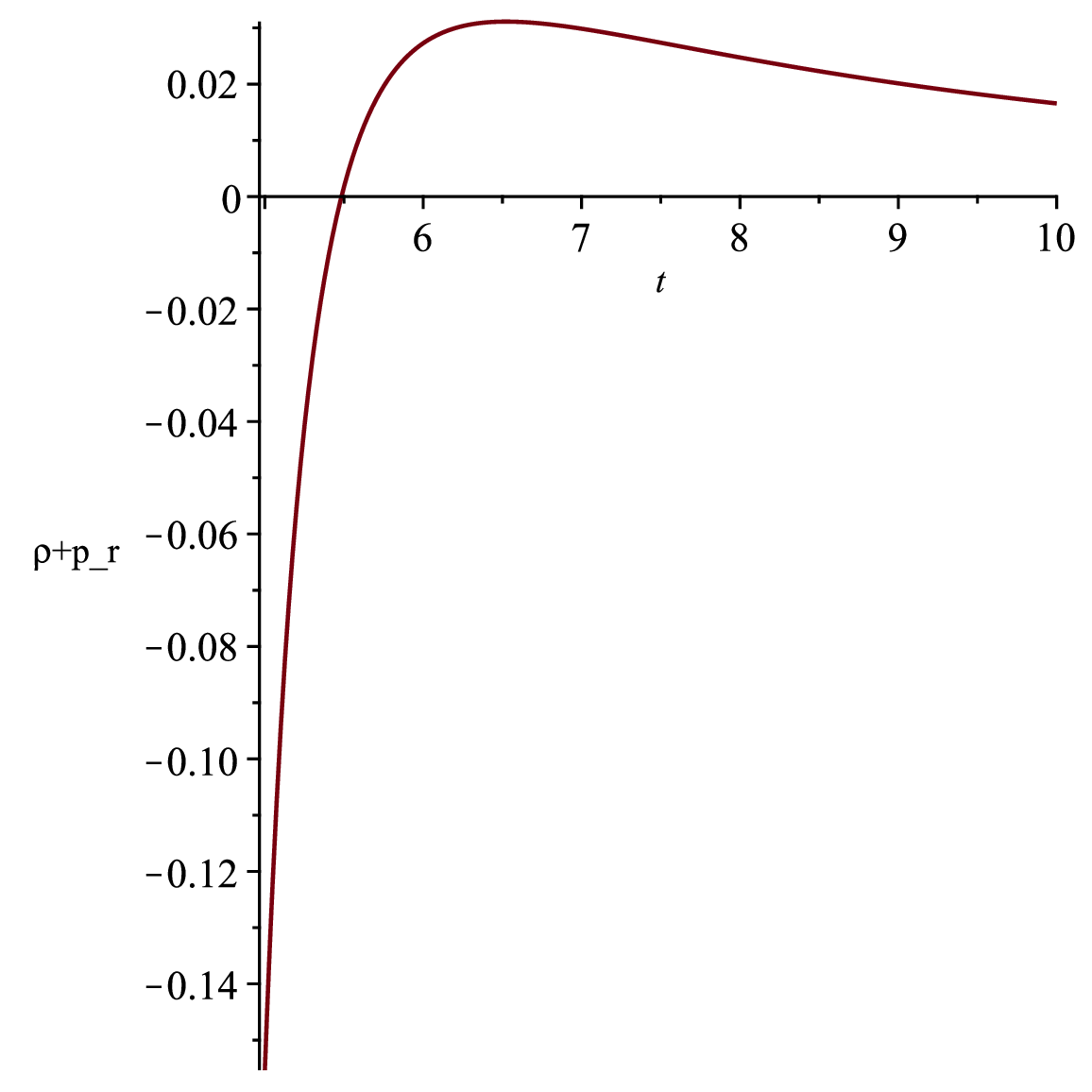}
	\caption { (left panel) In case 1,  during the evolution of the universe, when $t>t_0$, the NEC is obeyed  (middle  panel) For case 2, the wormhole is always supported by exotic matter   (right panel) In case 3.1,  during the evolution of the universe, when $t>t_0$,   again the NEC is  obeyed.}
\end{figure}

For case 1, equation (68) implies \[t> \left[ \frac{\delta -kr_0^2 }{nr_0^2}\right]^{\frac{1}{2n-2}}\equiv t_0,\]
for which  $(\rho + p_r)_{r=r_0} >0 $. Thus at the evolution of the universe, when $t>t_0$, the NEC is obeyed (see left panel of figure 2). But up to time $t<t_0$, we get wormhole supported by the matter violating NEC i.e. by exotic matter. Here, one can note that evolution of the universe plays a crucial role for the formation of wormhole.

For case 2, the wormhole is always supported by exotic matter. Here evolution of the universe does not affect on wormhole configuration(see middle panel of figure 2).\\

 For case 3.1, NEC obeys  for all time after \begin{equation} t > \left[\frac{ 72(\delta -kr_0^2) }{\beta^2r_0^2}+\frac{12(L+3k)}{\beta^2}\right]^{\frac{1}{2}} -C_1 \sqrt{\frac{3}{\beta}} =t_0.\end{equation}
But up to time $t<t_0$, the wormhole is supported  by exotic matter(see right panel of figure 2).\\

For case 3.2, NEC obeys if \begin{equation} \frac{1}{3} (\beta - L - 3k) >\frac{\delta -kr_0^2 }{r_0^2} .\end{equation}
Thus through out the evolution of the universe, one gets wormhole without violating NEC for the above condition (70)(see left panel of figure 3).\\

 For case 4, NEC obeys if \begin{equation}   t^{\alpha-2}e^{2 \gamma t}>\frac{\delta -kr_0^2 }{\alpha r_0^2} .\end{equation}
Hence, during the evolution of the universe, when $t>T$, where T is satisfying the equation
\[T^{\alpha-2}e^{2 \gamma T}-\frac{\delta -kr_0^2 }{\alpha r_0^2}=0,\]
the NEC is obeyed. But up to time $t<T$, the wormhole is supported  by exotic matter (see right panel of figure 3).\\
\begin{figure} [thbp]
\centering
	\includegraphics[width=6cm]{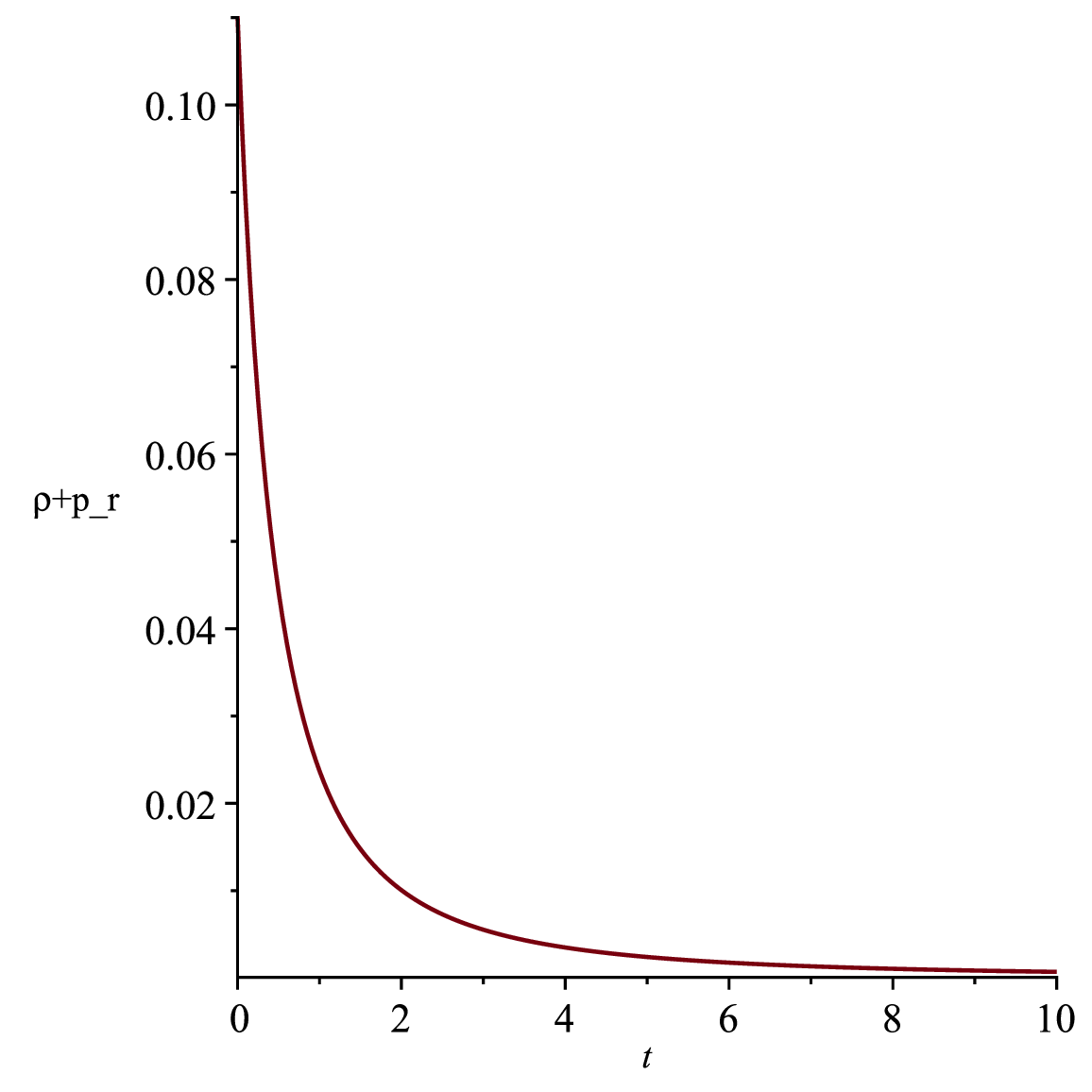}
\includegraphics[width=6cm]{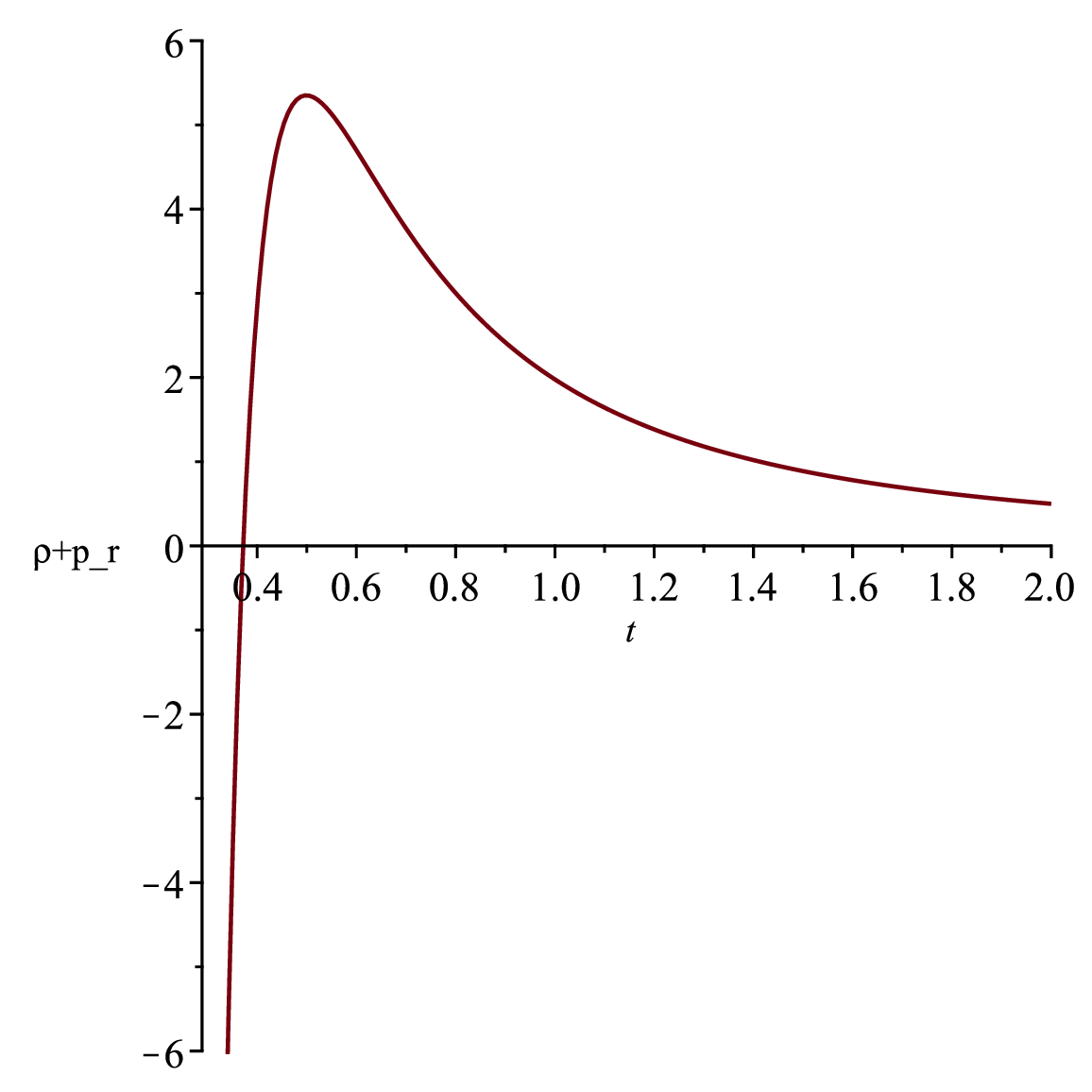}
	\caption { (left panel) In case 3.2,  through out the evolution of the universe, one gets wormhole without violating NEC   (right panel) In case 4, again the NEC is  obeyed after certain time.}
\end{figure}
The present spacetime metric is describing a dynamical wormhole in FRW expanding universe. Here, $a(t) \rightarrow \infty$ as $t\rightarrow \infty$. Also, all A for different proposed models except case 3.1 approach to zero as $t\rightarrow \infty$ and NEC is violated. So, NEC violations could be desisted only for finite non zero values of A. Thus we restrict finite interval of times in which $a(t)$  is finite. The time interval may arbitrary small or large where $a(t)$ is non zero finite in that interval. However, in the case 3.1, after the time $t=t_0$ given in equation (69) one can get a wormhole supported by non exotic matter for ever. This is possible because, as $t\rightarrow \infty$, A assumes non zero  finite   value.

\section{    traversability conditions:}

According to Morris - Thorne, for a convenient journey through the wormhole,
certain conditions should be imposed at the space stations ( initial and final destinations ) as well as on velocity of the traveler. The traveler feels the  acceleration  must  not exceed the   Earth's gravitational acceleration, $  g_\oplus$.
 One can calculate the traveler's four-acceleration
 in his proper reference frame as follows:
 The  orthonormal basis  $(e_{\widehat{0}}' , e_{\widehat{1}}', e_{\widehat{2}}' , e_{\widehat{3}}')$  of the traveler's proper
  reference is obtained via Lorentz transformation in
  terms of the orthonormal basis vectors  $(e_{\widehat{t}} , e_{\widehat{r}}, e_{\widehat{\theta}} , e_{\widehat{\varphi}})$ of  the static observers as
 \[ e_{\widehat{0}}' =  \sigma e_{\widehat{t}} \mp \sigma \frac{v}{c} e_{\widehat{r}} ,~ e_{\widehat{1}}' =
  \mp \sigma e_{\widehat{r}} + \sigma \frac{v}{c} e_{\widehat{t}}, ~e_{\widehat{2}}'  = e_{\widehat{\theta}} , ~e_{\widehat{3}}'  = e_{\widehat{\varphi}} ,\]
 where $\sigma  = (1-\frac{v^2}{c^2})^{- \frac{1}{2}}$ , and $v(r)$ be   the velocity of the traveler  when  he passes $r$ as observed   by a static observer
positioned there.

To travel through a wormhole, the tidal
gravitational forces experienced by a traveler must be reasonably
small. According to Morris and Thorne, the acceleration felt
by the traveler should not exceed Earth's gravity. Thus the tidal
accelerations between two parts  of the traveler's body,
separated by several  meters, must less than the gravitational
acceleration at Earth's surface $g_\oplus$ .
One can calculate the tidal acceleration felt by the
traveler as follows:

  The tidal acceleration   is given by the formula
  \[
\Delta a^{\hat{\mu^\prime}}=-R_{\hat{\nu^\prime}\hat{\alpha^\prime}\hat{\beta^\prime}}^{\hat{\mu^\prime}}U^{\hat{\nu^\prime}}\eta^{\hat{\alpha^\prime}}
U^{\hat{\beta^\prime}} ,\]
where $U^{\hat{\mu^\prime}} = \delta^{\hat{\mu^\prime}}_{\hat{0^\prime}} $  is
the traveler's four velocity and  $\eta^{\hat{\alpha^\prime}}$
is the separation between two arbitrary parts of his body   which
is purely spatial in the traveler's reference frame.
The nonzero components of Riemann  tensor  $R_{\hat{2^\prime}\hat{0^\prime}\hat{2^\prime}\hat{0^\prime}}$ in  the traveler's frame from the  static  observer's  frame  can be obtained by  using  a  Lorentz  transformation.

Now the radial tidal constraint is given by
\begin{equation}R_{\hat{1^\prime}\hat{0^\prime}\hat{1^\prime}\hat{0^\prime}}=R_{\hat{r}\hat{t}\hat{r}\hat{t}}=
     \left|\frac{\ddot a}{a} \right| \leq \frac{g_{\oplus}}{c^2 |\eta|}.
\end{equation}

Note that, for the evolving wormhole, the tidal acceleration depends on time. More specifically, in this expanding wormhole, the gravitational forces experienced by an observer at a constant throat radius decreases over time. In fact, the radial tidal constraint $(72)$ is directly restraining the expansion of the wormhole. If one considers the size of the traveler's body is $| \eta | = 2 $ m, then

\begin{equation*}
    \frac{g_\oplus}{c^2 |\eta|} \approx \frac{1}{(10 ^8 m)^2}.
\end{equation*}
\\
Now, we calculate the radial tidal acceleration constraint of the evolving wormholes obtained in this paper with the different scale factors and have found the restriction on time as follows:\\

$\textbf{Case 1:}$
\begin{equation}
    t \geq c \sqrt{n(n-1)}
\end{equation}
\\
For, a particular value of $n$, say $n=2$, $c = 10^8$ $m/sec$, we have
\begin{equation*}
    t \geq 1.414 \times 10^8 \, sec.
\end{equation*}

\textbf{Case 2:} \\

In this case of exponentially expanding wormhole universe, the radial tidal forces experienced by a traveler is independent of the time  evolution. It has a restriction on the parameter $\alpha$ as
\begin{equation}
    \omega \geq \frac{1}{ \sqrt{c} }.
\end{equation}
\\
\textbf{Case 3.1:}
\begin{equation}
    t \geq \left[ \left(   \frac{2 c^2 \beta}{3} - \frac{4(L+3k)}{\beta}  \right)^\frac{1}{2} - C_1 \right] \sqrt{\frac{3}{\beta}}.
\end{equation}

Here for the following values of the parameters, say $\beta = 1, k = 0, L = 0, C_1 = 0, c = 10^8 m/sec$, we have
\begin{equation*}
    t \geq 1.414 \times 10^8 \, sec.
\end{equation*}

\textbf{Case 3.2:}\\

In this case, $\ddot a = 0$, i.e. the radial tidal acceleration is identically zero and hence satisfies the constraint $(72)$ automatically.
\\

\textbf{Case 4:}
\begin{equation}
    t \geq \frac{2 \alpha \gamma c^2 + \sqrt{4 \alpha ^2 \gamma ^2  c^4 + 4 c^2 (\alpha ^2 - \alpha) ( 1 - \gamma ^2 c^2)}}{2 ( 1 - \gamma ^2 c^2 )}.
\end{equation}
For a particular choice of parameters, say $\gamma c = 0.2, \alpha = 3$ and $c = 10^8 \, m/sec$, we have the following restriction on time $t$,
\begin{equation*}
    t \geq 8.571 \times 10^7 \, sec.
\end{equation*}
Therefore, it is speculated that after a fixed time of the evolution, one may have a traversable wormhole. However, there is some lateral tidal constraint also.

The lateral tidal constraint is given by
\[
R_{\hat{2^\prime}\hat{0^\prime}\hat{2^\prime}\hat{0^\prime}}=R_{\hat{3^\prime}\hat{0^\prime}\hat{3^\prime}\hat{0^\prime}}=\sigma^2
R_{\hat{\theta}\hat{t}\hat{\theta}\hat{t}}+\sigma^2\frac{v^2}{c^2}  R_{\hat{\theta}\hat{r}\hat{\theta}\hat{r}}  + 2 \sigma^2\frac{v}{c} R_{\hat{\theta}\hat{t}\hat{\theta}\hat{r}}\]
\begin{equation}
   = \left| \sigma ^2 \frac{\ddot a}{a} - \frac{\sigma ^2 \frac{v^2}{c^2}}{2 a^2 r^3}\left[2 \dot a ^2 r ^3 - b - k r^3 + r(b' + 3 k r^2)\right]   \right| \leq \frac{g_\oplus}{c^2 |\eta|}.
\end{equation}
Lateral tidal constraint gives a restriction on velocity of the traveler as well as matter distribution comprising the wormhole. We have derived the limitation for the lateral tidal acceleration and get the restriction of non-relativistic velocity at the throat (i.e. $v<<c, \, \sigma \approx 1)$ as follows:\\

$\textbf{Case 1: }$
\begin{equation}
      v  \leq \sqrt{2} t^n \left[ c^2 n (n-1) t^{-2} +1  \right]^{\frac{1}{2}} \left[  2 n^2 t^{2n -2} + L + 2K + \rho _s e^{- \frac{r_0}{r_s}} - r_0 ^{-2}  \right]^{- \frac{1}{2}}.
\end{equation}
$\textbf{Case 2:}$
\begin{equation}
    v  \leq \sqrt{2} e^{\omega t} \left( \omega ^2 c^2 + 1  \right)^{\frac{1}{2}} \left[ 2 \omega ^2 e^{2 \omega t} + 2K + L + \rho _s e^{- \frac{r_0}{r_s}} - r_0 ^{-2}  \right]^{- \frac{1}{2}}.
\end{equation}
$\textbf{Case 3.1:}$
\begin{equation}
      v  \leq  \sqrt{2} \left[ \frac{1}{4} \left( t \sqrt{\frac{\beta}{3}} + C_1  \right)^2 + \frac{L + 3K}{\beta}  \right]^{\frac{1}{2}} . \left(  \frac{\beta}{6} c^2 + 1 \right)^{\frac{1}{2}} .\left[ \frac{\beta}{6} \left( t \sqrt{\frac{\beta}{3}} + C_1  \right)^2 + L + 2K  + \rho _s e^{- \frac{r_0}{r_s}} - r_0 ^{-2}    \right]^{- \frac{1}{2}}.
\end{equation}
$\textbf{Case 3.2:}$
\begin{equation}
     v  \leq  \sqrt{2} \left[ t \sqrt{\beta - (L + 3K)} + \sqrt{3} C_1 \right]^{\frac{1}{2}}    . \left[  2 \beta + L + 3 \rho _s e^{- \frac{r_0}{r_s}} - 3 r_0 ^{-2}  \right]^{- \frac{1}{2}} . 3^{\frac{1}{4}}.
\end{equation}
 $\textbf{Case 4: }$
\begin{equation}
      v \leq  \sqrt{2} \left( t^\alpha  e^{\beta t} \right)^{\frac{1}{2}} . \left[ c^2 \{ (\alpha ^2 - \alpha) t^{\alpha - 2} + 2 \alpha \beta t^{\alpha - 1} + \beta^2 t^{\alpha}  \} e^{\beta t} + 1  \right]^{\frac{1}{2}}  . \left[ 2 \left( \alpha t^{-1} + \beta \right)^2 e^{2 \beta t} t^{2 \alpha} + L + 2K +  \rho _s e^{- \frac{r_0}{r_s}} - r_0 ^{-2}    \right]^{- \frac{1}{2}}.
\end{equation}

$a=1$, $k=0$ gives the case for static wormhole.

\section{Proper Length Between Two Distances $r_1$, $r_2$:}
It's equally intriguing to investigate how changes in the scale factor over time impact the nature of the proper length or proper circumference of the throat of the wormhole.\\

Let us first consider the case of proper circumference of the throat of wormhole. Consider a slice of the wormhole
at a fixed instant of time and $\theta = \pi / 2$. Let $r_0$ be the throat radius of the wormhole so that $r = b(r_0) = r_0$. The proper circumference $(C_0)$ of the wormhole throat is
\begin{equation}
    C_0 = \int_{0} ^{2 \pi} \, a(t) r_0 \, d \phi = a(t) (2 \pi r_0).
\end{equation}
This is merely the scale factor times the static wormhole circumference.\\
Now we calculate the radial proper length $( l(t))$ between any two points $r_1 > r_0$ and $r_2$ as
\begin{equation}
      l(t)  = \pm a(t) \int_{r_1}^{r_2} \frac{dr}{(1 - k r^2 - \frac{b}{r})^{1/2}},
\end{equation}
which is just scale factor times of the radial proper separation of the static wormhole counterpart ( see figure 2).\\
\begin{figure} [thbp]
\centering
	\includegraphics[width=6cm]{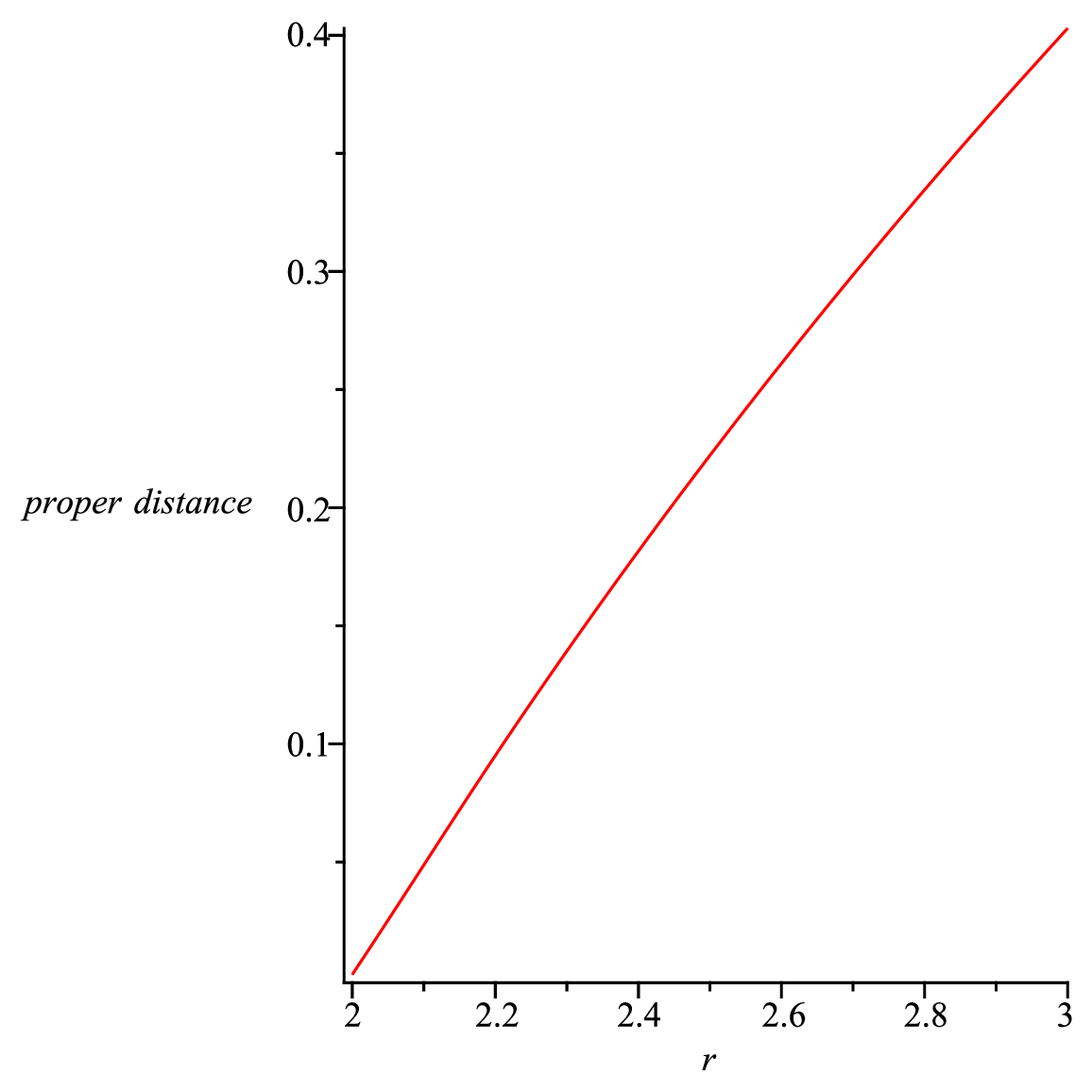}
	\caption {  The radial proper separation of the static wormhole counterpart.}
\end{figure}
However, a temporal singularity may occur at $t = t_0$ where $a(t_0) = 0$ .\\

\section{concluding remarks:}

Within this manuscript, we explore the viability of wormholes containing dark matter within the framework of classical general relativity against a backdrop of Friedmann-Robertson-Walker (FRW) cosmology. Our investigation extends to dynamic wormholes situated in various cosmological contexts, each defined by distinct criteria such as specific choices of scale factors and cosmological density. The temporal existence of these geometries, subject to the null energy condition (NEC) for the involved matter, introduces an intriguing aspect. Despite the potential perturbations associated with their transient nature, this dynamical perspective presents an advancement over static geometries. Notably, wormholes supported by non-exotic matter can be established within a finite interval of time when the scale factor $a(t)$ assumes a non-zero finite value. The duration of this evolution can span from arbitrarily small to large intervals. Section 3.1 highlights a scenario where the evolution period can be infinitely large. After the time $t = t_0$ (as given in equation 69), a wormhole supported by non-exotic matter can endure indefinitely. Our findings reveal instances where the proper circumference of the wormhole throat expands due to the temporal evolution of the geometry. Turning our attention to traversability criteria related to the tidal forces experienced by a traveler, we adopt a systematic approach. This involves transitioning from the static observer's frame in the case of a static wormhole to the comoving frame for the evolving geometry. Subsequently, we apply a simple Lorentz transformation to switch to the traveler's frame. Riemann tensor components are then derived in this traveler's frame, establishing constraints on tidal forces. These constraints necessitate determining the time at which the inequalities impose the most stringent conditions through extremization. Subsequent extremization of the traveler's velocity is required to obtain the ultimate condition. Satisfying these conditions renders the wormhole traversable. To quantify traversability, we determine the maximum speeds at which a traveler could traverse the wormhole, considering specific values of parameters and the throat radius $r_0$. This ensures compliance with the given constraint.

\subsection{Acknowledgements}
We are grateful to the referee for his constructive suggestions which help a substantial improvement of the manuscript.
FR    would like to thank the authorities of the Inter-University Centre for Astronomy and Astrophysics, Pune, India for providing the research facilities. We are also thankful to Somi Aktar for her help in checking the solutions. FR and  BSC  are also   thankful to SERB, DST  $\&$   DST FIST programme     (  SR/FST/MS-II/2021/101(C)) and   UGC, Govt. of India for financial support respectively.

\end{document}